\documentclass[%
 reprint,
 amsmath,amssymb,
 aps,
 prl,
]{revtex4-2}

\usepackage{amsmath}
\usepackage{acro}
\usepackage{subcaption}
\usepackage{graphicx}
\usepackage{dcolumn}
\usepackage{bm}
\usepackage{hyperref}
\usepackage{xcolor}
\usepackage{caption}

\DeclareAcronym{MLE}{short=MLE, long=Maximum Likelihood Estimate}

\newcommand{\vars}{\bm{\alpha}}
\newcommand{\hyperparams}{\bm{\beta}}
\newcommand{\x}{\mathbf{x}_k}
\newcommand{\y}{\mathbf{y}_k}
\newcommand{\modelsymbol}{y_m}

\newcommand{\modelk}{\boldsymbol{\modelsymbol}(\x;\vars)}
\newcommand{\modeli}{\modelsymbol(x_i,\vars)}
\newcommand{\md}{\boldsymbol{\delta}(\mathbf{x})}

\newcommand{\cov}{\boldsymbol{\Sigma}}
\newcommand{\covmd}{\boldsymbol{\Sigma}^{\text{md}}}
\newcommand{\datak}{\mathcal{D}_k}
\newcommand{\alldata}{\boldsymbol{\mathcal{D}}}
\newcommand{\fullcov}{\boldsymbol{\mathcal{V}}}
\newcommand{\delk}{\boldsymbol{\Delta}_k}

\NewDocumentCommand \suite {m} {\texttt{#1}}

\begin{document}


\title{The East Lansing Model: \\
a Bayesian uncertainty quantified optical potential for rare isotopes}
\author{K. Beyer}
\email{beyerk@frib.msu.edu}
\affiliation{Facility for Rare Isotope Beams, Michigan State University, East Lansing, Michigan 48824, USA}
\affiliation{Department of Physics and Astronomy, Michigan State University, East Lansing, Michigan 48824, USA}

\author{F. M. Nunes}
\affiliation{Facility for Rare Isotope Beams, Michigan State University, East Lansing, Michigan 48824, USA}
\affiliation{Department of Physics and Astronomy, Michigan State University, East Lansing, Michigan 48824, USA}


\date{\today}

\begin{abstract}
The East Lansing Model is a global, uncertainty-quantified optical potential for neutron and proton projectiles, with a novel form for the neutron-proton asymmetry component, with the goal to improve extrapolations away from stability. Our Bayesian calibration relies on (n,n), (p,p) and (p,n) experimental data for angular distributions on spherical targets with mass $A\geq 40$, and beam energies in the range $E = 10-100$ MeV. When considering the stable nuclei for which data is available, our results demonstrate that the inclusion of the $(p,n)$ data alone does not significantly change the parameterization. The additional information contained in (p,n) only becomes evident by introducing a new parameterization, one that gives the flexibility to encode neutron skins in the optical potential through an asymmetry dependent term. 
Finally, extrapolations of ELM toward the limits of stability (namely toward the proton and neutron driplines) leads to reduced uncertainties when compared to other global optical potentials in use.
\end{abstract}

\maketitle

{\it Introduction} One of the top quests in science is to identify where the heaviest elements on Earth came from. To answer this mind-capturing question, astrophysical modeling is making impressive strides. Simulations of large stars now include full 3D evolution \cite{freytag2023} as well as star rotation (e.g. \cite{choplin2018}). We now understand that roughly half of the abundance of the elements in the solar system is produced in explosive scenarios and involve isotopes far from stability. Astrophysical modeling of explosive events has also made incredible progress, including full 3D large-scale computations for supernovae  \cite{sn2025} and neutron star mergers \cite{nsm2024},  as well as the possibility of strong magnetic fields \cite{magnetic2023}. These astrophysical simulations involve a nuclear network that traces through a sequence of reactions with neutron-rich nuclei, nuclei that have never been measured in a laboratory before. The information contained in those nuclear networks mostly relies on theories fitted to stable nuclei, theories for which the uncertainties away from stability are unknown. This is the critical piece that the field is missing.

Fission is a well-known process with widespread societal applications, including energy production and defense \cite{schunck2022}. The fragments resulting from fission are themselves neutron-rich. Due to the existence of many neutrons in the environment, it is critical to understand how these neutron-rich fragments interact with other neutrons. Yet again, existing models extrapolate theories fitted on stable nuclei to regions outside their training, resulting in unknown uncertainties \cite{beyer2024uncertainty}.

Rare isotope facilities such as the Facility for Rare Isotope Beams can perform measurements to constrain nuclear processes away from stability. However, the fierce competition for beam time means that such measurements will be rare, and the field will continue to rely heavily on extrapolations from stability. The issue of extrapolating beyond the known regions is widespread in science. Although popular in the research community, data-driven methods, like neural networks and Gaussian processes, are problematic when used in extrapolations far from the trained region: either the uncertainty estimate is unrealistically small or it blows up \cite{hasson2020,cho2025}. Understanding (and encoding) the physical laws behind the data patterns is one of the best ways to perform controlled extrapolations.

In this work, we focus on nucleon scattering on a target. Although this is a complex many-body problem, the optical model projects it onto the ground state of the target, introducing a nucleon-target effective interaction -- the so-called optical potential. Optical potentials are ubiquitous in models for nuclear reactions, and are widely used in both astrophysics (for example in nucleosynthesis networks \cite{Hebborn_2023}) and in terrestrial applications \cite{beyer2024uncertainty}. Global optical potentials attempt to capture the systematics of the nucleon-target effective interaction as a function of mass, charge and beam energy. Often these global potentials depend on many dozens of parameters, which need to be constrained by data.

Thus, a critical aspect in extracting optical potentials from data is data curation. Global optical potentials have traditionally been fitted with elastic scattering observables, mostly angular distributions but also reaction (or total) cross sections, and polarization observables (e.g. Konig Delaroche (KD) \cite{KD} and the Chapel Hill (CH) \cite{CH} parameterizations).  Extrapolation to neutron-rich nuclei is governed by the dependence of the global optical potential neutron-proton asymmetry. While the systematics along an isotopic chain provide some information, we demonstrate that charge-exchange reactions, like (p,n), contain crucial information for extrapolating to neutron-rich systems.

The nuclear force acts similarly on neutrons and protons -- this is referred to as isospin symmetry, which allows for expressing the optical potential in terms of an isoscalar and an isovector component \cite{lane1962isobaric,lane1958r}, as done in the CH global parameterization \cite{CH}.  In fact, it has been shown that predictions of radiative neutron capture rates on exotic, neutron-rich isotopes relevant to r-process nucleosynthesis are strongly sensitive to the isospin dependence of the optical potential \cite{goriely2007isovector}. Although the original CH \cite{CH} only used elastic scattering to constrain the parameters, there have been isolated historical efforts to include $(p,n)$ reactions in the fitting, with mixed success \cite{lovas1976}. Those studies assume the same shape for both isoscalar and isovector components, an assumption questioned by a recent analysis of a couple of specific elastic and charge-exchange reactions \cite{danielewicz2017symmetry}. 

It was only recently that the first uncertainty-quantified global optical potentials were developed \cite{cole2023}. Based on the original global parameterizations of KD and CH, Pruitt et al. \cite{cole2023} performed a Bayesian calibration to determine KDUQ and CHUQ, respectively. The corresponding posterior distributions are well constrained for most of the parameters, with a notable exception for the isovector strength, which determines the dependence on neutron-proton asymmetry, and, therefore, the behavior of the potential when extrapolated. It has been shown that, when the uncertainties from KDUQ \cite{cole2023} are propagated to $(p,n)$ reactions populating the Isobaric Analog States (IAS), the credible intervals for the cross sections are extremely wide and, in some cases, the agreement with existing data is poor \cite{smith2024uncertainty,whitehead2022prediction}. 

The East Lansing Model (ELM) presented here is a global spherical optical potential with Bayesian uncertainty quantification that preserves isospin symmetry while allowing for the isovector and isoscalar terms to have independent radial profiles. It is a global potential, with $21$ parameters, constrained by all available elastic and (p,n) IAS data on spherical targets above mass $A=40$, data that spans beam energies $E=10-100$ MeV. Our study shows that the ELM framework greatly improves the reliability of extrapolations to neutron-rich systems when compared to KDUQ and CHUQ. 
\\

{\it Methods:} The functional form of the optical potentials is key for extrapolating away from stability, and should encode the isospin symmetry of the nuclear force. ELM borrows from CH  the Lane form \cite{lane1962isobaric}, which couples projectile and target isospins ($\boldsymbol{\tau}$ and $\boldsymbol{T}$ respectively). In the Lane form, there are two components, the isoscalar term that depends only on the target mass $A$, and the isovector term which controls the $(N-Z)/A$ dependence in elastic scattering:  $U(r) = U_0(r) + \frac{\boldsymbol{\tau} \cdot \boldsymbol{T}}{A} U_1(r)$. 

In fact, $(p,n)$ to the IAS is mediated by exactly this isovector term, and therefore provides an independent constraint on it from elastic scattering along isotopic chains, or neutron and proton scattering on the same target. Global optical potentials used today do not take advantage of the full power of this symmetry; most omit $(p,n)$ data (e.g. \cite{varner1991global}), and many produce separate parameterizations for $n$ and $p$ projectiles, ignoring isovector symmetry completely (e.g. \cite{koning2003local}). In addition, even when using the Lane form, global parameterizations assume that $U_0$ and $U_1$ have the same geometric form factor.


The East Lansing Model is different from previous global parameterizations in three different ways: (i) the shape of the interactions, (ii) the data used as evidence and (iii) the statistical model used for calibration and uncertainty quantification.

{(i) The shape of the optical potential:} The form for the real, central part of ELM is: 
\begin{equation}
    \label{eq:elm}
    V(r)  =  V_0 f_0(r) \pm \frac{N-Z}{A} V_1 f_1(r),
\end{equation}
\noindent
with $f_{0(1)}$ being isoscalar (isovector) Woods Saxon form factor, each with its own independent depth, radius $R_{0(1)}$ and difuseness $a_{0(1)}$ parameters. The isoscalar and isovector radii are scaled by mass through the relation $R_{0(1)} = r^{0}_{0(1)} + r_{0(1)} A^{1/3}$. 
In Eq.\ref{eq:elm}, the $+(-)$ sign refers to proton (neutron) projectiles. 

Given that CH is Lane consistent, we use the CH parameters as prior information in a Bayesian analysis. We also fix the spin-orbit term to CH, to avoid an ill-posed statistical problem. The full ELM has 21 free parameters. To understand the relevance of the independence in geometries of isoscalar and isovector components, we also perform a calibration on a simplified version with 18 parameters, called ELM0, which imposes the same radius and difuseness for the isoscalar and the isovector term. The full phenomenological parameterizations are provided in the supplemental material. 




{(ii) The data used as evidence:} A body of experimental differential cross sections between $10 < E < 100$ MeV were assembled from EXFOR \cite{data2022exfor} using \suite{exfor-tools} \cite{beyerexfortools}. The data set spanned 17 entries for $(p,n)$, with 710 total data points, and 57 entries for elastic scattering, comprised of 5965 data points. Targets considered where near-spherical with $0^+$ ground states, and only $(p,n)$ with well-resolved $\Delta J^\pi = 0^+$ transitions to the IAS where considered. The data curation workflow is reproducible, employing a set of open-source \suite{jupyter-notebook}s, linked in the supplemental material.

{(iii) The statistical model:} The Bayesian calibration of a physics model $\modelsymbol$ can be posed \cite{kennedy2001bayesian}:

\begin{align}
\begin{split}
    \underbrace{y_i}_{\text{observation}} + \underbrace{\epsilon_i}_{\text{uncertainty}} & \equiv \underbrace{\zeta(x_i)}_{\text{truth}} \\ 
    &= \underbrace{\modeli}_{\text{model}} + \underbrace{\delta(x_i)}_{\text{discrepancy}} 
\end{split}
    \label{eq:stat_model}
\end{align}
\noindent
in which the experimental data $y_i$ and model output (at a given point in input space, $x_i$, and for given set of parameters, $\vars$) $\modeli$, are fixed, but the rest of the variables are random. The goal is to simultaneously learn the posterior distribution of model parameters, and of the model discrepancy, to make uncertainty-quantified predictions of the latent truth $\zeta$ at new locations in the input space. 

For global optical models, the input space consists of mass and charge of the target $A,Z$, the observable class: $\{(p,p)_{\text{elastic}}, (n,n)_{\text{elastic}}, (p,n)_{\text{IAS}} \}$, the lab-frame bombarding energy $E$, and the center-of-mass frame scattering angle $\theta$. The experimental data consist of differential cross sections for many $A,Z,E$ combinations. We will use $k$ to label each angular distribution, $i$ to label $\theta$ within each data set, and $l$ to label the observable class. The set of experimental constraints can then be represented by $\boldsymbol{\mathcal{D}} \equiv \{ \{ \mathcal{D}^l_k \equiv \{ x_i, y_i \}_{i=0}^{N_k} \}_{k=0}^{M_l} \}_{l}$.

 Because systematic errors are reported inconsistently across the experimental literature, we assume that the experimental covariance can be approximated as diagonal:
$ \cov_k \approx \delta_{ij} \sigma_i^2$, where $\sigma_i$ is the reported statistical uncertainty of $y_i$. 
The model discrepancy is taken as a multivariate normal $\md \sim \mathcal{N}(0, \covmd)$, this time over the entire input space. 
In ELM, we use a parametric, diagonal model, following \cite{pruitt2023uncertainty}:   $ \covmd_{k \in l} \approx \delta_{ij}  \left( \beta_l \tilde{y}_i  \right)^2 $,
thus determining that, at each point $x_i$, the standard deviation of the model discrepancy is a fraction $\beta_l$ of $\tilde{y}_i$ (the average of the data and model output). The $\beta_l$  hyperparameters are inferred alongside $\vars$. These choices neglect correlations in either the experimental error or model discrepancy between points $x_i, x_j$.

Finally, the marginal likelihood for $\datak$ is:
\begin{equation}
    L(\datak|\vars,\beta_l) 
    = \left( (2\pi)^{N_k} |\fullcov| \right)^{-1/2} \exp{\left( \delk^\top \fullcov_k^{-1} \delk  \right)},
\end{equation}
\noindent
where the full covariance is $\fullcov_k \equiv \cov_k + \covmd_k(\beta_l)$ and the residuals are defined by $\delk \equiv \y - \modelk$. Then, given a prior $p(\vars)$ and a hyperprior $p(\hyperparams)$, we use Bayes' rule to determine the posterior, trivially combining independent data sets with equal weights:
\begin{align}
\label{eq:post}
    \begin{split}
        p(\vars,\hyperparams | \alldata) p(\alldata) 
            &= p(\alldata | \vars, \hyperparams) p(\vars) p(\hyperparams) \\
            &= \left[ \prod_k L(\datak|\vars,\hyperparams)  \right] p(\vars) p(\hyperparams).
    \end{split}
\end{align}
The evidence $p(\alldata)$ is independent of $\vars$ and $\hyperparams$, and is irrelevant for sampling.
Note that our statistical model also differs from that of \cite{cole2023}, most notably in the scaling of the log-likelihood. In \cite{cole2023}, the log-likelihood is scaled by the ratio of the number of free model parameters to the total number of experimental observations, effectively inferring a power posterior \cite{pmlr-v19-grunwald11a}, whereas we employ no rescaling, and infer the standard Bayesian posterior.

Sampling from Eq.~\eqref{eq:post} was done using Markov Chain Monte Carlo. The prior for $p(\vars)$ was chosen as a product of uncorrelated normal distributions, with means and widths determined from CHUQ \cite{pruitt2023uncertainty}. The priors for the new isovector geometry parameters in ELM where estimated from \cite{danielewicz2017symmetry}. The open-source hierarchal Bayesian model calibration library \suite{rxmc} \cite{rxmc} was used to evaluate the log-likelihood, which in turn used the open-source package \suite{jit$\mathcal{R}$} \cite{beyerjitr} to solve the Schr\"{o}dinger scattering equation with the calculable $\mathcal{R}$-matrix method \cite{descouvemont2010r} and compute the resulting elastic and $(p,n)$ cross sections. A description of the Markov Chain Monte Carlo implementation and numerical convergence of the calibration are discussed in the supplemental material.\\

\begin{figure*}[t]
    \centering
    \includegraphics[width=0.6\linewidth]{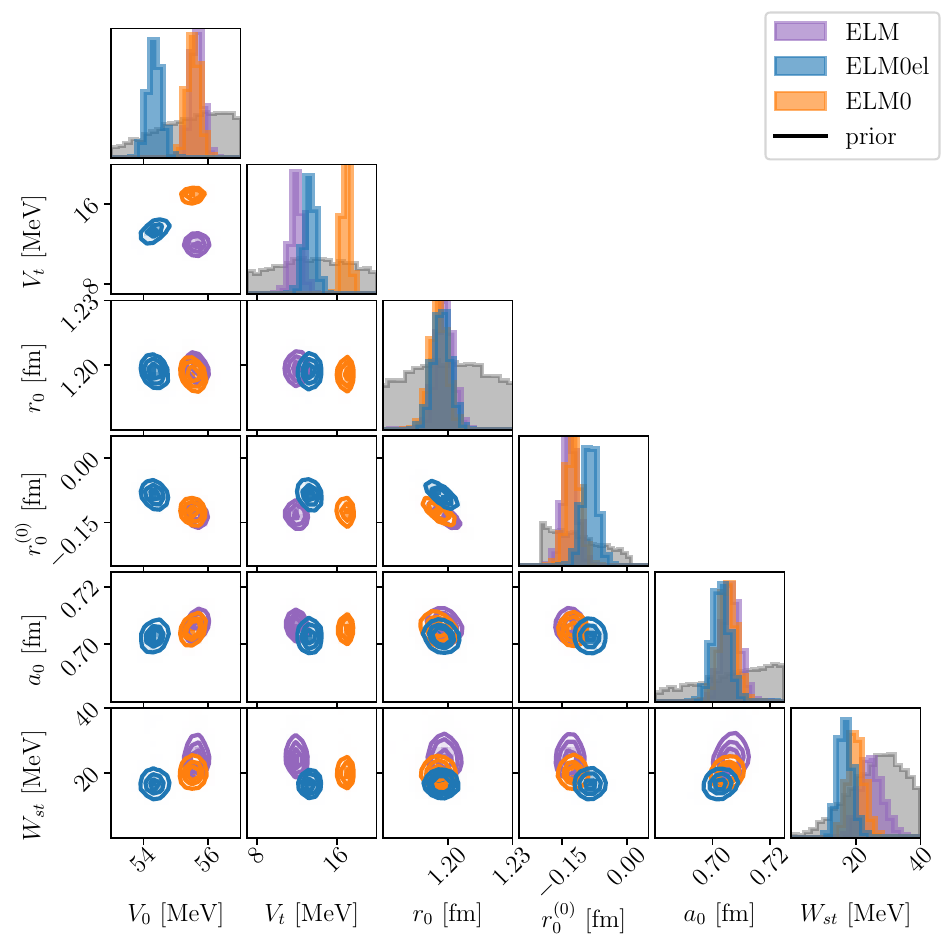}
    \caption{Corner plot of a subset of the model parameters, including isoscalar depth, radii parameters and difuseness ($V_0,r_0,r^{(0}_0, a_0$), and isovector real and imaginary ($V_t,W_{st}$). The ELM (purple), ELM0 (orange) and ELM0el (blue) are compared with the prior distributions used (grey). See supplemental material for detailed definitions.}
    \label{fig:corner}
\end{figure*}

\begin{figure}
    \centering
    \includegraphics[width=0.7\linewidth]{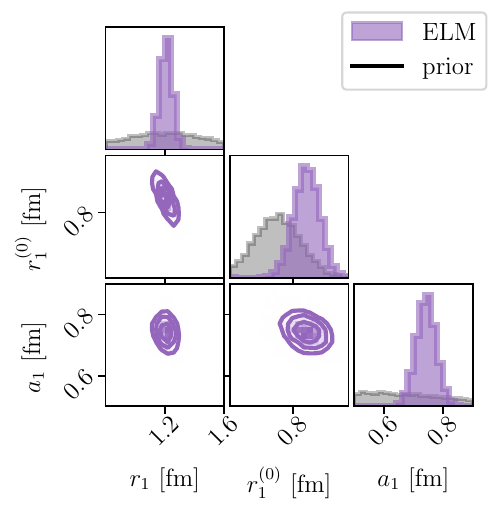}
    \caption{Corner plot for the isovector geometry parameters $r_1,r^{(0}_1, a_1$ unique to ELM (purple), compared to the prior (grey).}
    \label{fig:corner_geom}
\end{figure}

\begin{figure}
    \centering
    \includegraphics[width=0.7\linewidth]{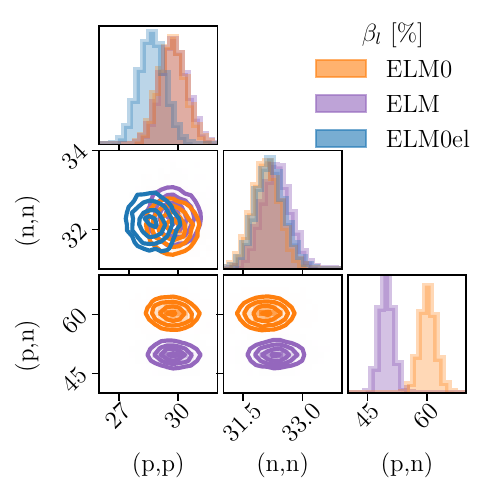}
    \caption{Corner plot for the fractional uncertainty parameters $\beta_l$ in the model discrepancy for each observable class $l$ \{$(p,p)$, $(n,n)$ or $(p,n)$\} expressed as a percentage, for ELM (purple), ELM0 (orange), ELM0el (blue). Note that ELM0el was not fit to $(p,n)$, and therefore has no model discrepancy for it.}
    \label{fig:corner_unc}
\end{figure}

{\it Results} The corner plot resulting from the Bayesian calibration of ELM is shown in  Fig.\ref{fig:corner}, for a selected set of parameters. Along the diagonal are the parameter posterior distributions, and the off-diagonal display the correlations. We compare three global calibrations. In purple is shown the full ELM calibration. In orange, ELM0 is shown, which is calibrated to the same data as ELM, but imposes the same geometry for isovector and isoscalar components. In blue is ELM0el, which has the same model form as ELM0, but omits the $(p,n)$ data in the calibration. ELM0el therefore corresponds to the standard approach to fitting a global, Lane-consistent optical potential to elastic scattering data (e.g. CH89 \cite{varner1991global}). By comparing blue with orange, it is clear that $(p,n)$ data forces the maximum likelihood to move in parameter space, significantly increasing the depth of the isovector term. However, this change is essentially corrected when the additional geometric parameters are introduced (comparing orange and purple). Both elements need to be introduced simultaneously. We found that the mean values obtained for ELM are significantly different than those of CHUQ.


We also performed a global calibration for ELMel, which assumes the same parameters as ELM but excludes $(p,n)$ data. This calibration failed to converge after half a million samples. Evidently, $(p,n)$ data is required to constrain the isovector geometry. 

In Fig \ref{fig:corner_geom}, we show the posterior distributions of the new parameters, unique to ELM, namely the isovector radii $A^{1/3}$-dependence $r_1$ and intercept $r_1^{(0)}$, and isovector diffuseness $a_1$. These can be compared to their isoscalar counterparts in Fig. \ref{fig:corner}. In ELM, the isovector geometry is slightly more diffuse than the isoscalar, and has roughly the same $A^{1/3}$ radial dependence as the isoscalar, with the exception that the difference between intercepts $r_1^{(0)} - r_0^{(0)}$ is $\sim$ 1 - 1.5 fm.

Figure \ref{fig:corner_unc} displays the posteriors for the fractional uncertainty hyperparameters $\beta_l$.
All three models describe the elastic data with similar fractional uncertainty, but ELM is able to achieve a roughly 10\% reduction in uncertainty over ELM0 in the $(p,n)$ sector. These fractional uncertainties are  consistent with those obtained in \cite{pruitt2023uncertainty}. 

Using Eq. 26 of \cite{danielewicz2017symmetry} to convert the isovector component of the optical potential into neutron skins, we find neutron skin predictions by ELM scale as $R_{skin}(^{A}Z) = x \cdot (N-Z)/A$, where $x$ is roughly 1.59 fm for $A \sim 48$, decreasing linearly to 1.35 fm for $A \sim 208$. The extraction of an uncertainty-quantified neutron skin thickness from the global analysis of elastic and $(p,n)$ data will be the focus of a future publication.

\begin{figure}[t!]
    \centering
    \includegraphics[width=0.7\linewidth]{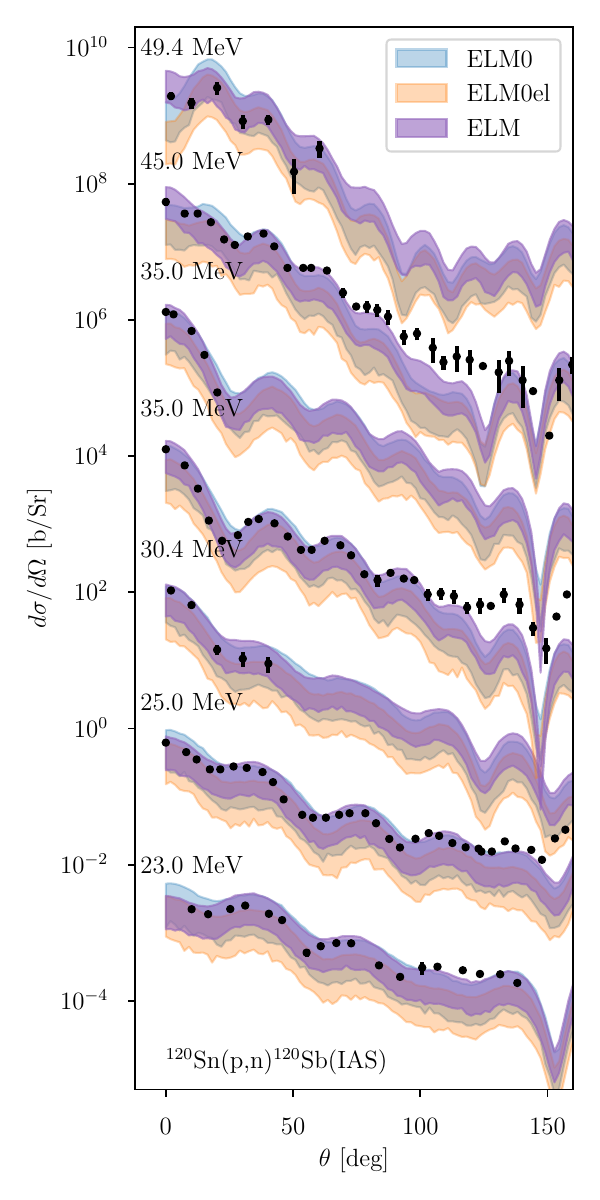}
    \caption{68\% credible intervals for the predictive posterior of the latent truth for $^{120}$Sn(p,n)$^{120}$Sb(IAS) from each model calibration, with experimental data in black: ELM (purple), ELM0 (orange), ELM0el (blue). Each successive energy is offset by a factor of 30 for visibility. }
    \label{fig:credible_interval}
\end{figure}

\begin{figure}
    \centering
    \includegraphics[width=\linewidth]{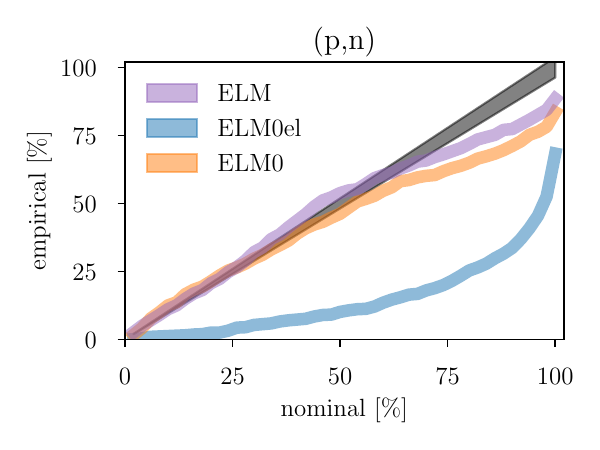}
    \caption{Nominal versus empirical coverage of the $(p,n)$ experimental data used for calibration: ELM (purple), ELM0 (orange), ELM0el (blue), compared with ideal coverage (grey). }
    \label{fig:ec}
\end{figure}   

\begin{figure}
    \centering
    \includegraphics[width=\linewidth]{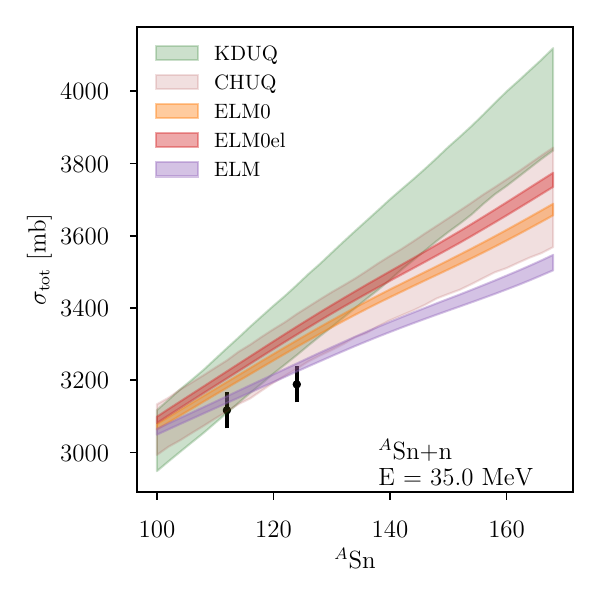}
    \caption{Neutron total cross sections on Sn isotopes from $A=$100 to 170:  68\% credible intervals predictions when extrapolating in asymmetry for the three models from this work -- ELM (purple), ELM0 (orange), ELM0el (blue) -- compared to CHUQ (pink) and KDUQ (green) from \cite{pruitt2023uncertainty}. Experimental data for $^{112}$Sn and $^{124}$Sn from \cite{pruitt2020isotopically}.}
    \label{fig:extrap_n}
\end{figure}

\begin{figure}
    \centering
    \includegraphics[width=\linewidth]{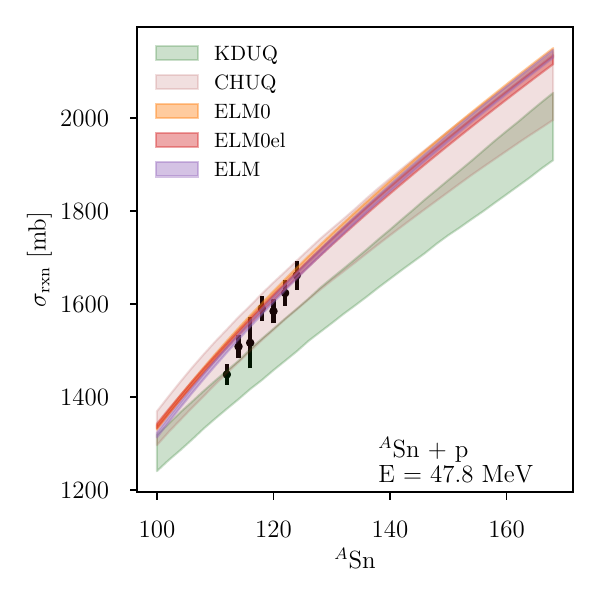}
    \caption{Proton reaction cross sections on Sn isotopes from $A=$100 to 170:  68\% credible intervals predictions when extrapolating in asymmetry for the three models from this work -- ELM (purple), ELM0 (orange), ELM0el (blue) -- compared to CHUQ (pink) and KDUQ (green) from \cite{pruitt2023uncertainty}. Experimental data is from \cite{carlson1995measurements}.}
    \label{fig:extrap_p}
\end{figure}



Next, we draw from the parameter posterior distributions to compute the $68$ \% credible intervals for the (n,n), (p,p) and (p,n) angular distributions. All models are equally able to describe the elastic scattering observables (not shown) but differ in their prediction for (p,n).  Fig.  \ref{fig:credible_interval} shows the angular distributions for $^{120}$Sn(p,n)$^{120}$Sb(IAS) at several beam energies, for the same three calibrations shown in Fig. \ref{fig:corner}, along with the experimental data (in black). 
Unsurprisingly, ELM0el, knowing nothing about (p,n) data,  fails to describe the $(p,n)$ angular distributions, while the inclusion of the independent isovector geometry along with $(p,n)$ data (ELM in purple) produces the best description.

The credibility of the uncertainties determined through the Bayesian analysis can be assessed by looking at the empirical coverage. These are shown in Fig. \ref{fig:ec}; for model uncertainties to be reliable the prediction should line up with the diagonal line. Although for elastic channels, all models have perfect empirical coverage (not shown), for (p,n) we find ELM offers significant improvement.

Note that our simplified model discrepancy form, which follows previous work and assumes no correlations in the input space, inhibits the calibration process. While ELM and ELM0  have nearly identical  coverages of the $(p,n)$ data (Fig. ~\ref{fig:ec}), the shapes of the differential $(p,n)$ cross sections they predict are qualitatively different. The model discrepancy 
regularizes the likelihood, inflating predictive posteriors at each point, which limits the power to discriminate between the performance of the two models. This issue should be alleviated by using a fully correlated model discrepancy.

It is  useful to compare ELM with the now-popular models CHUQ and KDUQ of \cite{cole2023} particularly away from stability.
In Figs. \ref{fig:extrap_n} and \ref{fig:extrap_p}, we compare the predictive posteriors for total cross section (for neutrons) and reaction cross sections (for protons) for the chain of Sn isotopes from the doubly closed shell $^{100}$Sn all the way to $^{170}$Sn. 
For comparison, posteriors for KDUQ and CHUQ are also displayed. 
Note that none of the ELM models were calibrated to any neutron total or proton reaction cross sections, and none of any of the models displayed, including KDUQ and CHUQ, included Sn isotopes outside of the range from $A=112$ to 124 in their experimental constraints. 
Because the uncorrelated model discrepancy form used here, and for CHUQ and KDUQ, is not amenable to extrapolation to new observables (it has no information about how model discrepancy changes as a function of $E,A,Z$, observable class, etc.), these figures compare only the predictive posteriors in the model predictions, $y_m$, not the latent truth $\zeta = y_m + \delta$. While the predictions for the reaction cross section are all consistent with the data, the neutron total cross section data clearly favors ELM over the other optical potentials.

The intermediate bombarding energies chosen for these comparisons reflect the region where ELM is most accurate, at higher ($E \gtrsim 90$) MeV and lower ($E \lesssim$ 10 MeV), KDUQ provides a better description of the Sn isotopes for which data is available (naturally, having a much larger energy range included in its constraints). Interestingly, proton reaction cross sections on Sn isotopes show little dependence on the inclusion of $(p,n)$ and the isovector geometry, while neutron total cross sections are significantly different in the three ELM models, a trend that holds at all energies.


{\it Conclusions} We have developed the East Lansing Model, a new uncertainty quantified global optical potential valid for nucleon scattering on target with mass $A>40$ and for beam energies in the range $E=10-100$ MeV, including (n,n), (p,p) and (p,n) data in a consistent framework. Along with the new global optical potentials, we also produced a new, open-source workflow that enables curating experimental constraints and performing Bayesian optical model calibration using elastic and charge-exchange data. Our statistical model follows the general statistical approach of \cite{kennedy2001bayesian} and includes model discrepancy. Past work done in Ref.\cite{pruitt2023uncertainty} are a special case of our statistical model.

The East Lansing Model expresses the interaction in terms of an isoscalar and an isovector term, with independent geometric form factors. The resulting calibration is physically robust and can extrapolate away from the valley of $\beta$-stability with unprecedented precision. We compared the full ELM with a simplified version (ELM0) that takes the same geometric form factors for the isoscalar and the isovector terms. And we perform an additional calibration, removing the (p,n) data (ELM0el). All these global potentials are able to describe elastic scattering. Naturally, adding (p,n) data to the calibration is necessary for a good description of charge exchange, although if no freedom is given to the geometric parameters it artificially forces the strength of the isovector term to increase. Including both (p,n) data and the independent geometry of isoscalar and isovector leads to an improved description of the data with reduced uncertainties. 

With ELM, the uncertainties for scattering away from stability (for example on neutron-deficient $^{100}$Sn and neutron-rich $^{132}$Sn) remain small. For these systems, significant differences are seen when comparing cross sections predicted by ELM with previous global potentials KDUQ and CHUQ. This arises from different choices made, not in the experimental constraints or physical model form, but in the statistical model; namely, the re-scaling of the likelihood. We will discuss this more deeply in a forthcoming paper.


Another salient point can be made about the model discrepancy: ELM uses the same model discrepancy form as CHUQ and KDUQ, which neglects correlations across the input space, and, therefore, may be innacurate when extrapolated to targets outside the set of experimental constraints. 
More sophisticated model discrepancy forms such as Gaussian Processes \cite{kennedy2001bayesian}, can offer a better assessment of the uncertainties, especially away from stability. 

Alongside to ELM, a new open-source model building platform, powered by \suite{jit$\mathcal{R}$}, \suite{rxmc} and \suite{exfor-tools}, has been implemented as part of the BAND framework \cite{bandframework}. As the statistical model is improved in the ways discussed above, as new EXFOR versions are released, and as phenomenological model forms are generalized (e.g. by coupling to low-lying rotational or vibrational states \cite{catacora2025}) or improved (e.g. via feedback from microscopic approaches \cite{baker2024}), this platform will be updated. We envision  improved versions of ELM to be released in the future, with the continual goal of robustly extrapolating towards the drip-lines\footnote{To use ELM, please visit \url{github.com/beykyle/elm} to obtain parameter samples, model definitions, and demonstrations of model calibration and propagation.}.



\textit{Acknowledgements}
This work was supported by the by U.S. Department of Energy grant DE-SC0021422, and the National Science Foundation CSSI program under award No. OAC-2004601 (BAND Collaboration \cite{bandframework}). This work relied on iCER and the High Performance Computing Center at Michigan State University for computational resources. 

The authors would like to thank C. D. Pruitt and P. Giuliani for valuable discussions.
\bibliography{b}

\end{document}